\documentclass[letterpaper,11pt]{article}
\pdfoutput=1

\usepackage{jheppub}
\usepackage{booktabs} 
\usepackage{graphicx} 
\usepackage{siunitx}
\usepackage{cleveref}
\usepackage{epsfig}
\usepackage[normalem]{ulem}
\usepackage{bm}
\usepackage{bbm}
\usepackage{slashed}
\usepackage{xspace}
\usepackage{subfigure}
\usepackage{multirow}

\usepackage{amsmath,latexsym}
\usepackage{pstricks}
\usepackage{color} 
 
\usepackage{marginnote}

\newcommand{\nn}{\nonumber}

\newcommand{\beq}{\begin{equation}}
\newcommand{\eeq}{\end{equation}}
\newcommand{\bea}{\begin{eqnarray}}
\newcommand{\eea}{\end{eqnarray}}

\begin{document}


\title{An Effective Field Theory of Quantum Mechanical Black Hole Horizons}

\author[a]{Walter D. Goldberger,}
\author[b]{Ira Z. Rothstein}
\affiliation[a]{Department of Physics, Yale University, New Haven, CT 06511}
\affiliation[b]{Department of Physics, Carnegie Mellon Univeristy, Pittsburgh, PA 15213}

\vspace{0.3cm}


\abstract{
We develop an effective theory which describes black holes with quantum mechanical horizons that is valid at scales long compared to the Schwarzschild radius but short compared to the lifetime of the black hole. Our formalism allows one to calculate the quantum mechanical effects in scattering processes involving black hole asymptotic states.
We point out that the EFT Wightman functions which describe Hawking radiation in the Unruh vacuum are not Planck suppressed and are actually {\it enhanced} relative to those in the Boulware vacuum, for which such radiation is absent. We elaborate on this point showing how the non-Planck suppressed effects of Hawking radiation cancel in classical observables.}
\maketitle



\section{Introduction}

The classical description of black holes is well understood, at least in principle.  Closed form solutions to
the equations of motion exist in the stationary cases but non-trivial dynamical solutions which include radiative effects are treated either
perturbatively or calculated numerically.  On the other hand, it is perhaps true that  a complete quantum mechanical description of black hole interactions is still lacking.  Of course, the quantum mechanical treatment of compact  macroscopic objects is typically
not a matter of interest due to decoherence. However, for black holes, the existence of the horizon implies   quantum mechanical effects, Hawking radiation \cite{hawking},  which must be treated in addition to those  of  collective coordinates. 
Hawking radiation of gravitons  can be regarded of as additional quantum gravity effects which must be
accounted for in considering dynamical process involving black holes. 

Quantum gravity on a fixed (smooth) background is understood \cite{DeWitt,tv} at sub-Planckian energies, and higher order corrections can be calculated 
within an effective field theory (EFT) framework \cite{donoghue}.  However, for black hole backgrounds, at present,
a formalism which allows for the calculation of all low energy quantum gravity effects is still lacking.
 This paper is an attempt to fill that void.  In particular, the  formalism introduced here has the power to calculate the Hawking radiation induced by a scattering event as well as the effects of Hawking radiation from scattering off black holes.

The  goal of this paper is to develop a formalism which allows for the calculation of quantum mechanical effects that arise in black hole (BH) spacetimes that is valid at distances large compared to the Schwarzschild radius. 
 However, we will restrict our calculations to times scales short compared to the lifetime of the black hole $\tau_{BH}\sim G_N^2 M^3$, effectively ignoring the back reaction due to the Hawking radiation.    We will always be in the regime in which the BH radius is super-Planckian, so that Hawking radiation remains a semi-classical process.   Thus we assume the following hierarchy of scales,
\beq
\tau_{BH} \gg \Delta \gg G_NM \gg 1/m_{Pl},
\eeq
where $\Delta$ is the typical length and time scale of the process under consideration and $m_{Pl}\sim G_N^{-1/2}$.

To properly include the effects of Hawking radiation within a worldline EFT \cite{GnR1,Goldberger:2006bd} we adapt the approach introduced in~\cite{GnR2} to account for the classical dissipation of long wavelength radiation at the black hole horizon.    In~\cite{GnR2}, such effects were attributed to the existence of gapless modes localized to the horizon which absorb energy as well as linear and angular momentum from the external environment.   These modes are presumably related to the fluctuations of the stretched horizon in the membrane paradigm \cite{Damour,Znajek,Price:1986yy}.    Regardless of their microscopic origin these modes can be ``integrated back in'' in order to systematically account for dissipative effects in the dynamics of black holes interacting with other compact objects. 

In the worldline approach, the state of these localized degrees of freedom corresponds to a ray in some Hilbert space $\cal{H}$.    In this description the semi-classical BH with mass $M\gg m_{Pl}$ then corresponds to a state $|M\rangle$ localized on the worldline.   In the absence of couplings to external (e.g. gravitational or electromagnetic) fields, the state $|M\rangle$ is an eigenstate of the BH Hamiltonian $H_0$.    The external fields couple to composite worldline operators which act on ${\cal H}$, thereby mediating transitions between the various eigenstates of $H_0$.   The correlation functions of these operators are obtained by a standard EFT matching calculation.  It is then possible to predict the effects of dissipation or emission in multi-body (eg scattering) processes in terms of the correlators of these operators, as done in~\cite{GnR2} for the case of classical absorption in BH/BH binary systems.

In this paper, which only introduces the formalism, we will consider a toy model of a quantized free real scalar field $\phi$ propagating in the BH background. Generalizing to photons or gravitons presents no formal obstruction to our methodology.   We write the leading order worldline coupling of the scalar field to the black hole as
\beq
\label{eq:EFT}
S_{BH} = -M\int d \tau -\int d\tau \phi(x(\tau)) {O}(\tau)+\cdots,
\eeq
where in this work, we neglect the effect of worldline operators that couple to higher partial waves of the bulk scalar than the $s$-wave.   The dynamics of the black hole horizon are then captured by the correlation functions of the operator $O( \tau)$.   We will then take $r_s\omega \ll 1$ as our power counting parameter, where $\omega$ is a typical frequency scale for the  process of interest.    The scaling of the operator $O$ with the power counting parameter will be determined by matching to the full theory of fields in the black hole background.

We will match to two different observables and show that they lead to the same worldline correlators.
First we will compare the EFT to the Wightman function $\langle 0|\phi(x)\phi(x')|0\rangle$ of the scalar field quantized in the Schwarzschild background, originally computed in~\cite{candelas}.    This is done in sec.~\ref{sec:candelas}, where we compute in the EFT the corrections to the Wightman function at spatial coincidence,
\beq
W_+(\omega,{\vec x})= \int dt  e^{i \omega t}\langle 0 | \phi(t,\vec x) \phi(0,\vec x) | 0 \rangle, 
\eeq
and, by comparison to the full theory result, extract the frequency space two-point Wightman function of $O(\tau)$,
\beq
\label{eq:Aplus}
A_+(\omega) = \int_{-\infty}^{\infty} d\tau e^{i\omega\tau} \langle M|O(\tau) O(0)|M\rangle.
\eeq

In sec.~\ref{sec:bekenstein}, we will perform an independent matching of the EFT to the transition probabilities $p(m\rightarrow n)$ calculated in \cite{wald,bekenstein}.    The quantities $p(m\rightarrow n)$ are defined as the probabilities that the black hole emits $n$ quanta of the field $\phi$ in a fixed mode given that there are $m$ incoming particles in the same mode.    These observables include the effects of both stimulated and spontaneous emission, and, as might be expected, yield more information about the EFT correlators than the two-point Wightman function considered in sec.~\ref{sec:candelas}.   In particular, based on the results of~\cite{wald,bekenstein}, we are able to show that to leading order in the power counting, the $n$-point Wightman functions of $O(\tau)$ are Gaussian, i.e. fully determined by the two-point correlator.

On the other hand the method presented in sec.~\ref{sec:candelas}, based on matching the full theory propagator, has the advantage that it provides an important consistency check of our formalism.   Namely, the results of~\cite{candelas} are presented for the three canonical choices of
boundary conditions, corresponding to the Boulware~\cite{boulware}, Unruh~\cite{unruh} and Hartle-Hawking~\cite{HH} states.    Because the retarded two-point function of an operator that satisfies linear field equations (e.g. a free-field)  is independent of the initial state, we must find that in the EFT, the retarded Green's function,
\beq
G_R(\tau) = -i\theta(\tau) \langle [O(\tau),O(0)]\rangle,
\eeq
must be the same regardless of the choice of state in the full theory.   We verify that this is the case by obtaining $A_+(\omega)$ when the full theory is either in the Boulware state, i.e. no Hawking emission from the BH, or in the Unruh state, corresponding to a BH radiating into empty space.    In the Boulware state, the response of the BH is purely absorptive, i.e. $A_+(\omega <0)=0$, corresponding to a classical black hole, while in the Unruh state the Wightman response is both dissipative and emissive.    Remarkably, the effects of  Hawking radiation, corresponding to non-vanishing $A_+(\omega <0)$, are not suppressed by powers of $1/m_{Pl}$ relative to classical absorption.     However, as we explicitly show in sec.~\ref{sec:retarded} to the next-to-leading order (NLO) in $r_s\omega \ll 1$, the effects of Hawking radiation cancel in the commutator $A_+(\omega)-A_+(-\omega)$, which is found to be the same wether the full theory is either in the Boulware or Unruh states.    The dispersive representation of the frequency space retarded Green's function,
\beq
\label{eq:disp}
G_R(\omega)= -i\int \frac{d\omega^\prime}{2\pi} \frac{A_+(\omega^\prime)-A_-(\omega^\prime)}{\omega-\omega^\prime-i \epsilon},
\eeq  
then implies that the quantum properties of the black hole hole are not accessible to observables which only measure the retarded response, in particular in astrophysical BH/BH binaries.


\section{Matching the bulk propagators}
\label{sec:candelas}

Our goal in this section is to extract the two-point correlation functions of the operator $O$ localized on the BH worldline.   We will focus on the Wightman function defined in Eq.~(\ref{eq:Aplus}).   The other correlators, i.e. retarded and
Feynman,  can be generated  via dispersion relations, as in Eq.~(\ref{eq:disp}).

The Wightman function $A_+(\omega)$ is extracted by matching the two-point functions $\langle\phi(x)\phi(x')\rangle$ in the EFT and in the full theory of the scalar field propagating in the Schwarzschild black hole background.   Note that the EFT is an interacting theory.    In particular, the BH, treated as a point source, interacts with the scalar both indirectly via graviton exchange (in the BH rest frame $-M\int d\tau = -M\int d\lambda\sqrt{1+h_{00}}$) and directly trough the coupling to the operator $O(\tau)$ as shown in Eq.~(\ref{eq:EFT}).    The former effects reproduce the scattering of the scalar by the BH's gravitational field while the $-\int d\tau O\phi$ coupling encodes the effects of absorption/emission by the BH horizon.

\subsection{The full theory}

In order to extract the correlator $A_+(\omega)$, we will take as input the two-point function of a quantized free scalar field $\phi(x)$ propagating in the background of a Schwarzschild BH.    This quantity was computed in~\cite{candelas} for the Boulware, Unruh, and Hartle-Hawking vacua, corresponding to the cases of no Hawking radiation, evaporating BH, and eternal BH in thermal equilibrium with a thermal bath, respectively.

From~\cite{candelas}, the Wightman propagator in the Boulware state is given by
\bea
\label{eq:boul}
\nn
\langle B|\phi(x)\phi(x')|B\rangle &=& \int_{-\infty}^\infty {d\omega \over 4\pi \omega} \theta(\omega)e^{-i\omega (t-t')}  \sum_{\ell m} Y_{\ell m}({\vec n}) Y^*_{\ell m}({\vec n'}) \left[{\stackrel{\rightarrow}{R}_{\ell}(\omega|r) \stackrel{\rightarrow}{R}^*_{\ell}(\omega|r') }\right.\\
& & \left.+ \stackrel{\leftarrow}{R}_{\ell}(\omega|r)  \stackrel{\leftarrow}{R}^*_{\ell}(\omega|r')\right],
\eea
while 
\bea
\nn
\langle U|\phi(x)\phi(x')|U\rangle &=& \int_{-\infty}^\infty {d\omega \over 4\pi \omega} e^{-i\omega (t-t')}  \sum_{\ell m} Y_{\ell m}({\vec n}) Y^*_{\ell m}({\vec n'}) \left[ {\stackrel{\rightarrow}{R}_{\ell}(\omega|r) \stackrel{\rightarrow}{R}^*_{\ell}(\omega|r') \over 1- e^{-\beta_H \omega}}\right.\\
& & \left.+\theta(\omega) \stackrel{\leftarrow}{R}_{\ell}(\omega|r)  \stackrel{\leftarrow}{R}^*_{\ell}(\omega|r')\right],
\eea
and
\beq
\langle H|\phi(x)\phi(x')|H\rangle = \int_{-\infty}^\infty {d\omega \over 4\pi \omega} e^{-i\omega (t-t')} 
\sum_{\ell m} Y_{\ell m}({\vec n}) Y^*_{\ell m}({\vec n'})\left[  {\stackrel{\rightarrow}{R}_{\ell}(\omega|r) \stackrel{\rightarrow}{R}^*_{\ell}(\omega|r') + {\stackrel{\leftarrow}{R}_{\ell}(\omega|r) \stackrel{\leftarrow}{R}^*_{\ell}(\omega|r')}\over 1- e^{-\beta_H \omega}}\right],
\eeq
are the Green's functions in the Unruh and Hartle-Hawking vacua respectively\footnote{ Our conventions for the definition of the Green's function differs from Candelas by a factor $i$.}.    In these expressions, the radial mode functions $\stackrel{\rightarrow}{R}_{\ell}(\omega|r)$, $\stackrel{\leftarrow}{R}_{\ell}(\omega|r)$ are solutions of the spin-zero Regge-Wheeler equation with boundary conditions
\beq
\stackrel{\leftarrow}{R}_{\ell}(\omega|r) = {1\over r}\left\{
\begin{array}{cc}
B_\ell e^{i\omega r^*}, & r\rightarrow r_s\\
e^{-i\omega r^*} +  \stackrel{\leftarrow}{A}_\ell e^{i\omega r^*}, & r\rightarrow\infty
\end{array}
\right.
\eeq
\beq
\stackrel{\rightarrow}{R}_{\ell}(\omega|r) = {1\over r}\left\{
\begin{array}{cc}
e^{-i\omega r^*} + \stackrel{\rightarrow}{A}_\ell e^{i\omega r^*},& r\rightarrow r_s\\
B_\ell e^{i\omega r^*},  & r\rightarrow\infty
\end{array}
\right.
\eeq
The radial tortoise coordinate is defined as $r^*=r+r_s\ln|r/r_s-1|$, and the coefficients $A_\ell$, $B_\ell$ (whose explicit frequency dependence is suppressed) satisfy the unitarity condition
\beq
|\stackrel{\rightarrow}{A}_\ell|^2 + |B_\ell|^2 = |\stackrel{\leftarrow}{A}_\ell|^2 + |B_\ell|^2 =1. 
\eeq
In light of this equation, we denote $|\stackrel{\rightarrow}{A}_\ell|$ and $|\stackrel{\leftarrow}{A}_\ell|$ by the common symbol $|A_\ell|$.    The greybody factor $|B_\ell|^2$ (transmission coefficient) is the same as the quantity $\Gamma_{\ell,m,\omega}$     referred to as the  ``absorption probability" in \cite{page}.   For spin $s=0$, $|B_\ell|^2~\sim (r_s|\omega|)^{2\ell + 2}$ for $r_s\omega\ll 1$.    The flat space limit corresponds to $B_{\ell}=0$ and $|A_\ell|=1$ (more precisely, given the asymptotic form of the flat space radial function, $j_\ell(z\rightarrow\infty)\sim \sin(z-\ell\pi/2)/z$, we see that $\stackrel{\leftarrow}{A}_\ell = (-1)^{\ell+1}$ in flat space.   The step function $\theta(\omega)$ in the Boulware state propagator Eq.~(\ref{eq:boul}) indicates that there is no Hawking radiation, and thus the response is that of a purely classical BH.   On the other hand, in the Unruh  and Hartle-Hawking states, the Wightman functions contain  positive and negative frequency parts so that there is both absorption and emission of radiation.

In performing the matching calculation, it is convenient to take the events $x,x'$ to be spatially coincident  $(r,\theta,\phi)$ at $r\rightarrow\infty$.   Using the asymptotic formulae~\cite{candelas} as $r\rightarrow\infty$,
\bea
\sum_{\ell=0}^\infty (2\ell +1) |\stackrel{\leftarrow}{R}_{\ell}(\omega|r)|^2 &=& 4\omega^2 +\cdots\\
 \sum_{\ell=0}^\infty (2\ell +1) |\stackrel{\rightarrow}{R}_{\ell}(\omega|r)|^2 &=& {1\over r^2} \sum_{\ell=0}^\infty (2\ell +1) |B_\ell(\omega)|^2+\cdots
\eea
the correlators at spatial coincidence and $r\rightarrow\infty$ take the form for $\Psi=B,U,H$
\beq
\label{eq:cunruh}
\langle \Psi|\phi(t,\vec x)\phi(t',\vec x)|\Psi\rangle = \langle\phi(t)\phi(t')\rangle_\Psi + {1\over 4\pi r^2} \int_{-\infty}^{\infty}  {d\omega\over 2\pi} e^{-i\omega (t-t')} \sum_{\ell} (2\ell+1) F_\ell^\Psi(\omega),
\eeq
which can be interpreted as the response function for an ``Unruh detector'' placed far away from the BH.   For $\Psi=B,U$ the function $\langle\phi(t)\phi(t')\rangle_\Psi$ is the flat spacetime scalar Wightman function at ${\vec x}={\vec x}'$,
\beq
\label{eq:fwight}
 \langle\phi(t)\phi(t')\rangle_{\Psi=B,U} \equiv \langle 0|\phi(t,\vec x)\phi(t^\prime,\vec x)|0\rangle = \int_0^\infty {\omega d\omega \over {4 \pi^2}} e^{-i\omega (t-t')},
\eeq
while for $\Psi=H$, it is the thermal Wightman function at the Hawking temperature $T_H=1/(4\pi r_s)$,
\beq
 \langle\phi(t)\phi(t')\rangle_{\Psi=H}\equiv {\mbox{Tr}\left[e^{-\beta H}\phi(t,\vec x) \phi(t^\prime, \vec x)\right]\over \mbox{Tr}[e^{-\beta H}]} = \int_{-\infty} ^\infty {\omega d\omega \over {4 \pi^2}} e^{-i\omega (t-t')}  {1\over 1- e^{-\beta\omega}}.
\eeq
The remaining term in Eq.~(\ref{eq:cunruh}) proportional to $1/r^2$ represents the effects of spacetime curvature and is given by
\bea
F^B_\ell(\omega) = {1\over 2\omega} \theta(\omega) {|B_\ell(\omega)|^2},
\eea
in the Boulware state, and 
\bea
\label{full}
F^U_\ell(\omega) = F^H(\omega)= {1\over 2\omega}  {|B_\ell(\omega)|^2 \over 1-e^{-\beta_H \omega}}
\eea
for the Unruh and Hartle-Hawking vacua respectively.

\subsection{Leading order EFT calculation}

The Wightman function $\langle \Psi|\phi(x)\phi(x')|\Psi\rangle$ above is an expectation value defined in the initial state of the quantum field around the BH background.   Because we are dealing with expectation values in a known initial state rather than transition matrix elements between fixed states in the far past and future, the observable to match in the EFT is the ``in-in" correlation function
\beq
G(x,x')=\langle in|\phi(x) \phi(x')| in\rangle.
\eeq
We will focus on the case in which the full theory is either in the Boulware or Unruh states, $\Psi=B,U$.   Given the form of the full theory result for $r\rightarrow\infty$, we take the initial state  $|in\rangle$ in the EFT to be 
\beq
\rho_{in} = |0\rangle\langle0|\otimes \rho_{BH},
\eeq
where $|0\rangle$ is the usual (Poincare invariant) free field vacuum, and $\rho_{BH}$ is some density matrix acting on the Hilbert space of BH states, whose form we need not specify in our calculation.

\begin{figure}[t]
    \centering
    \includegraphics[scale=0.3]{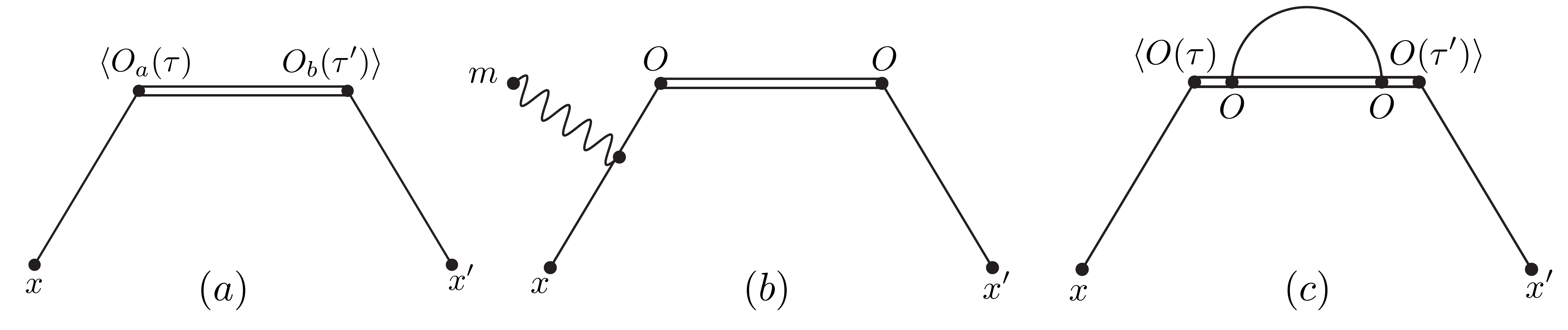}
\caption{Corrections to the Wightman function $\langle \phi(x) \phi(x')\rangle$ in the EFT.   Diagram (a) is leading order. Diagrams (b), (c) are next-to-leading order in the power counting.}
\label{fig:candelas}
\end{figure}

To compute this Wightman function we use the closed time path (CTP), or Schwinger-Keldysh, formalism \cite{CTP}.    We assume that the scalar field is minimally coupled to the static gravitational field of the point source at the origin.   In addition, $\phi$ couples to the source via the worldline interaction of Eq.~(\ref{eq:EFT}).   To leading non-trivial order in the power counting we have,
\bea
G(x,x') &=& W_0(x-x') +  \int d\tau d\tau' D^{2a}(x,x(\tau))  \langle O_a(\tau) O_b(\tau')\rangle D^{1b}(x(\tau'),x')+\cdots . 
\label{eq:inin}
\eea
The second term corresponds to Fig.~\ref{fig:candelas}(a), where $x(\tau)=(\tau,0)$ is the worldline of the black hole at rest at the origin of the coordinate system,  $D^{ab}(x,x')$ is the Schwinger-Keldysh propagator matrix of the free scalar
\beq
\label{eq:props}
D^{ab}(x,x') = \left(
\begin{array}{cc}
D_F(x-x') & - W_0(x'-x) \\
-W_0(x-x') & D_D(x-x')
\end{array}\right),
\eeq
with $D_F(x-x')=\langle 0|T\phi(x)\phi(x')|0\rangle$, $W_0(x-x')=\langle 0|\phi(x)\phi(x')|0\rangle$,  $D_D(x-x')=D_F(x-x')^*$ the interaction picture (i.e. free) propagators.   Explicitly, we have 
\bea
\label{eq:fprop}
D_F(x-x') &=& -{1\over 4 \pi^2} {1\over (x-x')^2 - i \epsilon} = -i\int_{-\infty}^{\infty} {d\omega\over 2\pi}  e^{-i\omega (t-t')} {e^{i|\omega| |{\vec x}-{\vec x}
|}\over 4\pi |{\vec x}-{\vec x}
|}, \\
 W_0(x-x') &=&  -{1\over 4 \pi^2} {1\over (x_0-x_0'-i\epsilon)^2 - ({\vec x} - {\vec x}')^2} = \int_{0}^{\infty} {d\omega\over 2\pi}  e^{-i\omega (t-t')}{\sin{\omega|{\vec x}-{\vec x}
|}\over 2\pi |{\vec x}-{\vec x}|}.
\eea
The matrix of worldline correlators appearing in Eq.~(\ref{eq:inin}) is defined as
\beq
\label{eq:omatrix}
 \langle O_a(\tau) O_b(\tau')\rangle  =  \left(
\begin{array}{cc}
\langle T O(\tau) O(\tau')\rangle & \langle O(\tau') O(\tau)\rangle \\
 \langle O(\tau) O(\tau')\rangle& \langle {\tilde T} O(\tau) O(\tau')\rangle 
\end{array}\right),
\eeq
where ${\tilde T}$ denotes the anti time-ordered product of operators.   In order to match the full theory at higher orders in powers of $r_s\omega,$ one would have to include diagrams where the scalar scatters off the worldline through graviton exchange (from expanding out $S_{BH}=-M\int dt\sqrt{1+h_{00}}+\cdots$), more insertions of ${ O}$, and interactions with higher multipole worldline operators (not displayed in Eq.~(\ref{eq:EFT})).   See Figs.~\ref{fig:candelas}(b),(c).   We will revisit these corrections in sec.~\ref{sec:retarded}.

Eq.~(\ref{eq:inin}) simplifies in the $r\rightarrow\infty$ limit, in which case one can drop rapidly oscillating phases $e^{\pm i \omega r}$ at the level of the integrand over frequency $\omega$.    The terms that remain can be expressed as 
\bea
\label{eq:glo}
G(t,\vec x; t', \vec x) &=& W_0(x-x') + {1\over (4\pi r)^2}  \int {d\omega\over 2\pi} e^{-i\omega (t-t')}\left[A_+(\omega) + \theta(\omega) (A_+(\omega) - A_+(-\omega))\right]\nn \\
\eea
where $A_+(\omega)$, defined in Eq.~(\ref{eq:Aplus}) is the frequency space Wightman function.   By comparing with the full theory state propagator Eq.~(\ref{eq:cunruh}), we therefore find that in the EFT, the  state $\Psi$ is described by a worldline theory whose two-point correlator is 
\beq
\label{eq:matching}
\langle  O(t) O(t')\rangle\equiv A^\Psi_+(t-t') = \int {d\omega \over 2\pi} e^{-i\omega(t-t')} A^\Psi_+(\omega),
\eeq    
where, for $\Psi=B$, 
\beq
\label{eq:ab}
A^B_+(\omega) \approx \theta(\omega) \omega \sigma^{abs}_{\ell=0}(\omega)= \theta(\omega) 4\pi r_s^2 \omega+\cdots.
\eeq 
Here we have used the relation $\sigma^{abs}_{\ell=0}(\omega) = {\pi\over\omega^2} |B_{\ell=0}|^2\approx 4 \pi r_s^2$~\cite{page} between the classical absorption cross section and the transmission coefficient.   Similarly, in the Unruh state\footnote{Technically, the relation between $A^{B,U}_+(\omega)$ and $\sigma_{abs}(\omega)$ in Eqs.~(\ref{eq:ab}),~(\ref{eq:au}) is only valid to leading order in the power counting.}
\beq
\label{eq:au}
A^U_+(\omega) = {\omega\sigma_{abs}(|\omega|)\over e^{\beta_H\omega}-1}\left[ 2 e^{\beta_H\omega}\theta(-\omega) + \theta(\omega) (1+e^{\beta_H\omega})\right]\approx 2\beta_H^{-1} \sigma_{abs}(|\omega|) = 2 r_s+{\cal O}(r_s\omega)^2,
\eeq
where we have dropped terms that are higher order in $r_s\omega \ll 1$.

As a check of these results, consider the flux of radiation seen by an observer at $r\rightarrow\infty$ from the black hole.   In the EFT, this is given by
\beq
\langle in| T^{rt}(x)|in\rangle = -{1\over 2} \lim_{x'\rightarrow x} \left(\partial_r \partial_{t'} +\partial_t\partial_{r'}\right) G(x,x').
\eeq
At $r\rightarrow\infty$, this reduces to
\beq
\langle in| T^{rt}(x)|in\rangle ={1\over 8\pi^2 r^2} \int {d\omega\over 2\pi} \theta(\omega) \omega^2  A_+(-\omega),
\eeq
from which we read off the differential energy emission rate from the BH
\beq
{d^2\over dt d\omega} M = {\omega^2\over 4\pi^2}   \theta(\omega)  A_+(-\omega)
\eeq
In particular, there is no energy flux in the Boulware state with, $A^B_+(\omega<0)=0$, while in the Unruh state the energy emission rate is $d^2 M/dt d\omega \approx r_s\omega^2/(2\pi^2)$, which matches the full theory Hawking emission spectrum~\cite{page} in the regime $r_s\omega\ll 1$.

\section{Matching Transition Probabilities}
\label{sec:bekenstein}
\subsection{The full theory}

Bekenstein and Meisels \cite{bekenstein} used thermodynamic arguments to derive a formula that yields the probability that a black hole emits $n$ identical spin-0 particles given that $m$ particles are incident in the same state in the far past.   Shortly after this, Wald and Panaganden  \cite{wald} verified the results in \cite{bekenstein} by calculating the $S$-matrix for a scalar field that propagates in the BH background but is otherwise non-interacting.   If one takes all the incoming and outgoing modes to be in the same normalizable wavepacket $|\psi\rangle$ of definite angular momentum $\ell$ which is sharply localized around some energy $\omega$, the transition probabilities read
\beq
\label{eq:bek}
p_{\ell} (m\rightarrow n) = {(1-x) x^n (1-|R_{\ell }|^2)^{n+m}\over (1 - x |R_{\ell }|^2)^{n+m+1}}\sum_{k=0}^{\mbox{min}(n,m)} {(n+m-k)!\over k! (n-k)! (m-k)! } \left[{(|R_{\ell }|^2-x) (1-x |R_{\ell }|^2) \over x (1-|R_{\ell }|^2)^2}\right]^k,
\eeq
where $|R_{\ell }(\omega)|$ is the reflection coefficient (with $|R_{\ell }(\omega)|^2=1-|B_{\ell}(\omega)|^2$) and for a non-rotating black hole, $x=\exp[-\beta_H \omega]$.

\subsection{The EFT Calculation}
To compare to the result of~\cite{bekenstein,wald}, we compute in the EFT the amplitude 
\bea
i{\cal A}(m+M\rightarrow n + X)=\langle X; n| T \exp\left[-i\int dt O(t) \phi(x)\right]|M;m\rangle
\eea
in the rest frame of the BH, taking an the initial state of the field $\phi(x)$ to be the $m$-particle state
\beq
|m\rangle = {1\over\sqrt{m!}} \left[\int {d^3 {\vec k}\over (2\pi)^3 2|{\vec k}|} \psi({\vec k}) a^\dagger(k)\right]^m |0\rangle.
\eeq
The state is normalized as $\langle m|m\rangle =1$, so that the wavepacket $\psi({\vec k})$ obeys the normalization condition $\int {d^3 {\vec k}\over (2\pi)^3 2|{\vec k}|} |\psi({\vec k})|^2=1$.    Similarly, the final state consists of $n$ particles in the same wavepacket.   Since the monopole operator $O(t)$ only couples to $s$-wave states, we take $\psi({\vec k})$ to be isotropic, of the form
\beq
\psi({\vec k}) = \sqrt{2\pi\over |{\vec k}|} \psi_0(|{\vec k}|),
\eeq
normalized according to $\int_0^\infty {dk\over 2\pi}|\psi_0(k)|^2=1$.    We assume that the function $\psi_0(k)$ is sharply localized around some frequency $\omega >0$.    

The transition probabilities are then 
\bea
p(m\rightarrow n) = \sum_X |{\cal A}(m+M\rightarrow n + X)|^2.
\eea
For example, the probability to absorb a single particle in the initial state is given by
\beq
p(1\rightarrow 0) \approx \sum_X \left|\int dt \langle X;0| O(t) |M\rangle \langle 0| \phi(x)|1\rangle\right|^2 ={\omega\over 2\pi} A_+(\omega).
\eeq
 Similarly, the single-particle emission probability is  $p(0\rightarrow 1)\approx {\omega\over 2\pi} A_+(-\omega)$.    On the other hand, from Eq.~(\ref{eq:bek}), we have in the full theory
 \beq
 p(1\rightarrow 0) \approx |B_0(\omega)|^2/\beta_H\omega+{\cal O}(r_s\omega)^2,
 \eeq
 and $p(0\rightarrow 1) = e^{\beta_H\omega} p(1\rightarrow 0) \approx  |B_0(\omega)|^2/\beta_H\omega+{\cal O}(r_s \omega)^2$.    Given that~\cite{page} $|B_0(\omega)|^2 = {\omega^2\over\pi} \sigma^{abs}_{\ell=0}(\omega)\approx  4 r_s \omega^2$, we therefore obtain, for $\omega>0$,
 \beq
 A_+(\omega)=A_+(-\omega) = 2 r_s.
 \eeq
 to leading order in the power counting.   This is in agreement with the results of sec.~\ref{sec:candelas} where we extracted the EFT worldline two-point functions by matching to the full theory propagator in the Unruh vacuum.

\subsubsection{Forward transition probabilities and IR divergences}

Unlike the full theory results computed in~\cite{wald,bekenstein}, the EFT transition probabilities suffer from infrared (IR) divergences.   In this section we provide a physical interpretation for such IR effects in the context of processes of the form  $n\rightarrow n$ for which such divergences arise at leading non-trivial order in the worldline interaction.

\begin{figure}
    \centering
    \includegraphics[scale=0.3]{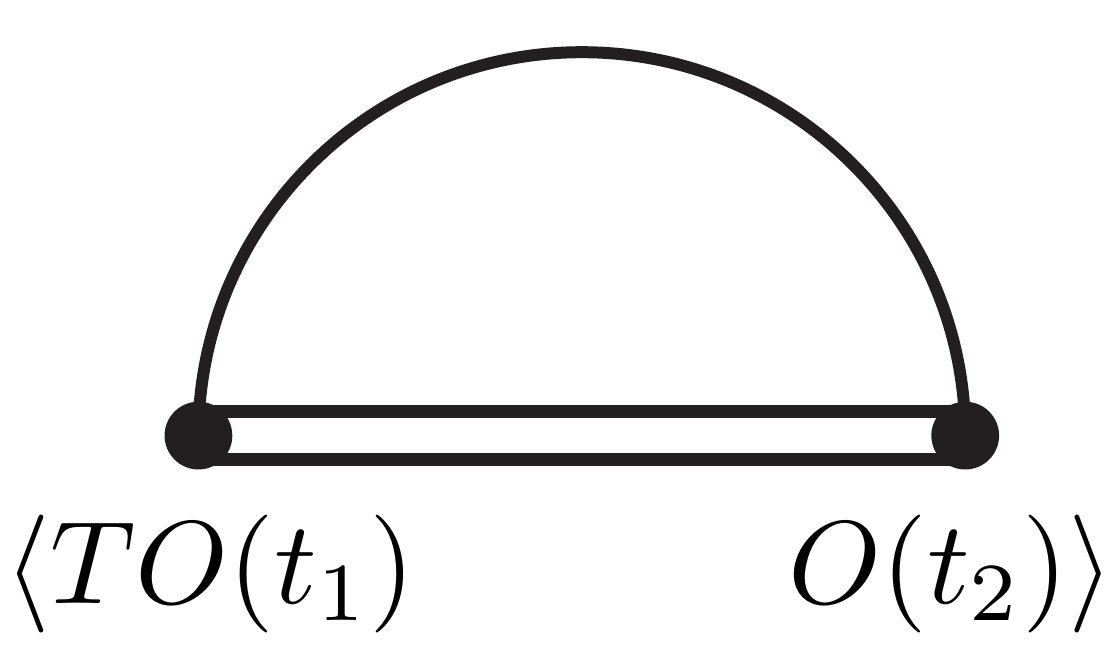}
\caption{Vacuum bubble contribution to the persistence amplitude}
\label{fig:vacbubble}
\end{figure}

First, consider the vacuum persistence probability $p(0\rightarrow 0)$, whose amplitude is to leading order
\beq
i{\cal A}(0+M\rightarrow 0 + X) \approx \langle X;0| 1 -{1\over 2!} \int dt_1 dt_2 T[\phi(x_1) O(t_1) \phi(x_2) O(t_2)]|M;0\rangle+\cdots.
\eeq
Squaring and summing over final states $X$, the first non-trivial term is due to interference, so that only the state $X=M$ appears in the final state sum.    Thus the vacuum-to-vacuum probability at lowest non-trivial order corresponds to the vacuum bubble diagram of Fig.~\ref{fig:vacbubble},
\beq
p(0\rightarrow 0) \approx 1 -{1\over 2} \int dt_1 dt_2 \left[D_F(x_1-x_2) \langle M|T O(t_1) O(t_2)|M\rangle+ \mbox{c.c}\right].
\eeq
By the results given above, we have to leading order in the power counting $\langle O(t) O(t')\rangle = 2 r_s \delta(t-t')$, and thus $\langle T O(t) O(t')\rangle = \langle {\tilde T} O(t) O(t')\rangle = 2 r_s \delta(t-t')$.   We can therefore write this equation as
\beq
p(0\rightarrow 0) \approx 1 - \int dt_1 dt_2 \,\mbox{Re}D_F(x_1-x_2) \, \langle M|T O(t_1) O(t_2)|M\rangle.
\eeq 
Finally, using Eq.~(\ref{eq:fprop}) for ${\vec x}_1={\vec x}_2=0,$
\beq
\mbox{Re}D_F(x_1-x_2)  = \int_{-\infty}^{\infty} {d\omega\over 2\pi} e^{-i\omega (t_1-t_2)} {|\omega|\over 4\pi},
\eeq
$p(0\rightarrow 0)$ takes the form
\beq
p(0\rightarrow 0)\approx 1 - {T} \int_0^\infty {d\omega\over 2\pi} {|\omega|\over 4\pi} \left(A_+(\omega) + A_+(-\omega)\right)+\cdots.
\eeq

This result is formally IR divergent, proportional to a time interval $T=(2\pi)\delta(\omega=0)\rightarrow\infty$.    This infrared divergence reflects a breakdown of the EFT at long times. In the full theory this divergence would get cut off by the lifetime of the black hole.   We can interpret this linear in time divergence as the first of an infinite set of singular terms involving higher powers of $T$, which we expect to sum up to $p(0\rightarrow 0) = \exp[-\Gamma T]$, where $\Gamma$ is the total decay width of the black hole.   Thus we identify
\beq
\label{eq:gamma}
\Gamma =\int_0^\infty {d\omega\over 2\pi} {|\omega|\over 4\pi} \left(A_+(\omega) + A_+(-\omega)\right),
\eeq
with the total decay rate of the black hole via Hawking emission of scalars.    Given that $A_+(\omega)\approx 2 r_s$, the integral over $\omega$ is UV divergent in the region $\omega\gg 1/r_s$ where the EFT breaks down.   However, the differential decay rate of the state $|M\rangle$,
\beq
{d\Gamma\over d\omega} = {\omega\over 8\pi^2} (A_+(\omega)+A_+(-\omega))\approx {r_s\omega\over 2\pi^2},
\eeq
is calculable and agrees with the low energy limit of Hawking's result \cite{hawking}.

Because $p(0\rightarrow 0)$ depends on an arbitrary IR time scale $T$, it is not a calculable quantity in the EFT.    The same IR divergence appears to leading non-trivial order in $p(n\rightarrow n)$,
\beq
p(n\rightarrow n) \approx 1-{1\over 2} \int dt_1 dt_2 \left[\langle n|T\phi(0,t_1) \phi(0,t_2)|n\rangle \langle M|T O(t_1) O(t_2)|M\rangle+ \mbox{c.c}\right],
\eeq
since by Wick's theorem, the correlator $\langle n| T\phi(0,t_1) \phi(0,t_2)|n\rangle= D_F(t_1-t_2) + \langle n|:\phi(0,t_1) \phi(0,t_2):|n\rangle$ breaks up into a contribution identical to the vacuum bubble in Fig.~\ref{fig:vacbubble} and an IR finite term associated with absorption followed stimulated emission of radiation,
\beq
p(n\rightarrow n) \approx p(0\rightarrow 0)-   \int dt_1 dt_2 \langle n|:\phi(0,t_1) \phi(0,t_2):|n\rangle \langle O(t_1) O(t_2)\rangle.
\eeq
We see that even though the forward probability $p(n\rightarrow n)$ is not calculable, the ratio $p(n\rightarrow n)/p(0\rightarrow 0)$
\beq
{p(n\rightarrow n)\over p(0\rightarrow 0)} \approx 1 -{n} \left({\omega\over 2\pi}\right) \left[A_+(\omega) + A_+(-\omega)\right]
\eeq
is IR finite.    This cancellation of IR divergences  that results from  normalizing $p(n \rightarrow n)$ by $p(0 \rightarrow 0)$ is expected to hold at higher orders in $\omega$ as in the usual linked cluster expansion.   Furthermore, compared to the low energy limit of the result in Eq.~(\ref{eq:bek})   
\beq
{p(n\rightarrow n)\over p(0\rightarrow 0)} \approx 1 - {2n} {|B_0(\omega)|^2\over \beta_H\omega},
\eeq
we find perfect agreement, providing another check of our formalism.

\subsubsection{Multi-particle probabilities and Gaussianity of EFT correlators}
\label{sec:gauss}

To extract the higher-point functions of the EFT operators, we need to match  transition probabilities with an uneven number  of initial and final external states. The amplitude takes the form
\bea
\nn
i{\cal A}(m+M\rightarrow n + X) &\approx& {(-i)^k\over k!} \int dt_1 \cdots dt_k \langle X|T O(t_1) \cdots O(t_k)|M\rangle \langle n |:\phi(x_1) \cdots \phi(x_k) : |m\rangle,\\
\eea
with $m\neq n$ and $k=|n-m|$.  Given that,
\bea
\langle n |:\phi(x_1) \cdots \phi(x_k) : |m\rangle\
=\left\{
\begin{array}{cc}
 \sqrt{n!\over m!} \psi^*(t_1)\cdots \psi^*(t_k), & (n>m)\\
 \sqrt{m!\over n!} \psi(t_1)\cdots \psi(t_k), & (n<m)
\end{array}
\right. 
\eea
where $\psi(t) = \sqrt{\omega\over 2\pi} \int_0^\infty {dk\over 2\pi} e^{-i k t} \psi_0(k)$, we find that after summing over the final black hole final states $X$, the transition probabilities take the form, for $n>m$,
\bea
{p(m\rightarrow n)} \approx {n!\over k!^2 m!} \left[\prod_{j=1}^k \int d{\tilde t}_j dt_j \psi({\tilde t}_j)\psi^*(t_j)\right]G^{(k,k)}({\tilde t}_1,\cdots {\tilde t}_k; t_1,\cdots t_k), & (n>m),
\eea
where the in-in Green's functions are
\beq
\label{eq:gnm}
G^{(n,m)}({\tilde t}_1,\cdots, {\tilde t}_n;t_1,\cdots, t_m) = \langle M|{\tilde T}[ O({\tilde t}_1) \cdots O({\tilde t}_n)] T[O(t_1) O(t_m)]|M\rangle.
\eeq
Similarly, for $n<m,$
\bea
{p(m\rightarrow n)} \approx  {m!\over k!^2 n!} \left[\prod_{j=1}^k \int d{\tilde t}_j dt_j \psi^*({\tilde t}_j)\psi(t_j)\right]G^{(k,k)}({\tilde t}_1,\cdots {\tilde t}_k; t_1,\cdots t_k), & (n<m).
\eea

On the other hand, the combinatorial factors that result from expanding the full theory Eq.~(\ref{eq:bek}) in the low frequency limit,
\beq
{p(m\rightarrow n)\over p(0\rightarrow 0)} \approx {(n+m-k_*)! \over k_*! (n-k_*)! (m-k_*)!} \left[{|B_0(\omega)|^2\over \beta_H\omega}\right]^{n+m-k_*}
\eeq
where $k_*=\mbox{min}(m,n)$, suggests that the correlators in the EFT are Gaussian, i.e. they factorize into products of two-point functions.   This is also suggested by the fact that the full theory consists of a free field propagating in a curved background spacetime.   We therefore make the ansatz that the $n>2$-point correlators in Eq.~(\ref{eq:gnm}) factorize into products over 2-point functions that depend on the branch of the closed time path, i.e $G^{(n,m)}$ follows from the generating functional
\beq
Z[J]=\exp\left[{1\over 2}\int d\tau d\tau' J_a(\tau) \langle O_a(\tau) O_b(\tau')\rangle J_b(\tau')\right],
\eeq
where $\langle O_a(\tau) O_b(\tau')\rangle$ was defined in Eq.~(\ref{eq:omatrix}).   A similar Gaussian ansatz for the correlators that describe classical scalar absorption by black holes was also recently considered in~\cite{Wong:2019yoc}.

For instance, consider the transition probability $p(0\rightarrow 2)$, which in the EFT is related to the 4-point function $G^{(2,2)}({\tilde t}_1, {\tilde t}_2;t_1,t_2)$.   Assuming that the correlators are Gaussian, this factorizes into products involving time-ordered, anti- time-ordered and Wightman two-point functions
\bea
\nn
G^{(2,2)}({\tilde t}_1, {\tilde t}_2;t_1,t_2) &=& \langle O({\tilde t}_1) O({t}_1)\rangle \langle O({\tilde t}_2) O(t_2)\rangle + \langle O({\tilde t}_1) O(t_2)\rangle \langle O({\tilde t}_2) O(t_1)\rangle\\
& & + \langle {\tilde T}O({\tilde t}_1) O({\tilde t}_2) \rangle \langle T O(t_1) O(t_2)\rangle.
\eea
The last term, involving $T$- and $\tilde T$-ordered products, cannot contribute to the transition probability since there are no particles in the final state, i.e. it vanishes by energy conservation. Thus for the case under study we are left with 
\bea
{p(0\rightarrow 2)} \approx {1\over 2!} \times 2 \left[\left({\omega\over 2\pi}\right) A_+(-\omega)\right]^2 = \left[p(0\rightarrow 1)\right]^2, 
\eea 
consistent with the low energy limit of the full theory.

More generally, assuming that Wick's theorem can be used to calculate the in-in correlators, we find that in the EFT the $n\rightarrow m$ transition probability is given by terms in which the contractions link together operators on opposite sides of the closed time contour.   There are $k!$ such terms, all of which contribute equally to the transition probability, so we find
\bea
{p(m\rightarrow n)\over p(0\rightarrow 0)} \approx \left(\begin{array}{cc} n\\ m\end{array}\right) \left[{\omega\over 2\pi} A_+(-\omega)\right]^{n-m} & (n>m),
\eea 
and the mirrored ($n \leftrightarrow m$) results holds when $m>n$.   Equivalently, the EFT prediction is
\beq
{p(m\rightarrow n)\over p(0\rightarrow 0)} \approx {(n+m-k_*)! \over k_*! (n-k_*)! (m-k_*)!} \left[{\omega\over 2\pi} A_+(-\omega)\right]^{n-k_*}  \left[{\omega\over 2\pi} A_+(\omega)\right]^{m-k_*}
\eeq
which is also consistent with the full theory, given that to leading order in the EFT, ${\omega\over 2\pi} A_+(\omega) = {\omega\over 2\pi} A_+(-\omega)=|B_0(\omega)|^2/\beta_H\omega$.

\section{The classical response function}
\label{sec:retarded}

It may seem paradoxical that the EFT Wightman response function $A^U_+(\omega)$, corresponding to the case where the full theory includes Hawking radiation, is not Planck suppressed relative to the function $A^B_+(\omega)$ extracted from the Boulware vacuum.   Rather we find that $A^U_+(\omega)$ is enhanced relative to $A^B_+(\omega)$ by one power of $(\beta_H\omega)^{-1}=1/(4\pi r_s\omega)\gg 1$.    So it would seem that Hawking radiation, which is a manifestly quantum effect, is not parametrically suppressed relative to the purely classical (absorptive) processes which the Boulware state describes.

The enhancement of $A^U_+(\omega)$ over  $A^B_+(\omega)$ has a simple interpretation as the Bose enhancement of the black body distribution as $T_H\rightarrow 0$, which is equivalent to the limit $r_s\omega \ll 1$ in which our EFT description is valid.   Despite this enhancement at the level of the Wightman functions, Hawking radiation contributes to ``classical observables" effects that are suppressed by powers of $\omega/m_{Pl}$ rather than $\beta_H\omega$, as one would intuitively expect.   In the full theory, this can be understood by considering the retarded propagator for the field $\phi$,
\beq
G_R(x,x^\prime)= -i\theta(t-t^\prime)\langle \Psi| [\phi(x),\phi(x')]|\Psi\rangle,
\eeq
which is the only correlator that is observable in classical BH processes, e.g. radiation from macroscopic binary systems.    In fact, the commutator $[\phi(x),\phi(x')]|$ is simply a $c$-number up to terms suppressed suppressed by $\omega/m_{Pl}$.   This follows because the full theory result in ref.~\cite{candelas} describes a free field propagating in a background gravitational field.   Neglecting the interactions, $\phi$ obeys linear equations of motion, so canonical quantization in the fixed background implies that $[\phi(x),\phi(x')]$ is proportional to the identity operator times a $c$-number function of $x,x'$.    (It is straightforward to check explicitly from the expressions in sec.~\ref{sec:candelas} that $\langle \Psi| [\phi(x),\phi(x')]|\Psi\rangle$  does not depend on $\Psi$).    Thus the effects of Hawking radiation on the retarded response function are suppressed by powers of the interactions.   In particular, for the more realistic case of gravitons, such self-interactions are suppressed by powers of $\omega/m_{Pl}\ll 1$, which are unobservable in classical processes.

On the other hand, in the EFT we have at coincident spatial points
\beq
\langle  [\phi(x),\phi(x')]\rangle  = {1\over 8\pi r^2}\int {d\omega\over 2\pi} e^{-i\omega t} \left[A^\Psi_+(\omega) - A^\Psi_+(-\omega)\right],
\eeq
as $r\rightarrow\infty$, so the statement that the retarded correlator in the full theory is state independent gets translated into a consistency relation on the EFT, namely
\beq
\label{eq:cr}
A^B_+(\omega) - A^B_+(-\omega) = A^U_+(\omega) - A^U_+(-\omega)  .
\eeq
To leading order in the power counting, we have from Eq.~(\ref{eq:ab}) that $A^B_+(\omega) - A^B_+(-\omega)\approx 4\pi r^2_s\omega$, while in the Unruh state $A^U_+(\omega) - A^U_+(-\omega) = 0 + {\cal O}(r_s\omega)$.   Due to the Bose enhancement of Hawking radiation in the Unruh state, verifying the consistency relation Eq.~(\ref{eq:cr}) requires a calculation at next-to-leading order in the power counting parameter.    This provides a non-trivial test of the EFT formalism, to which we turn to below.

\subsection{NLO matching in the Unruh state}

There are two types of corrections to the Wightman function in the EFT.   One is from corrections due to the gravitational potential interaction between the point source and the scalar field.   The second is from multiple insertions of the $-\int d\tau \phi {O}$ coupling in Eq.~(\ref{eq:EFT}).   Insertions of higher multipole operators, coupled to spatial gradients of $\phi$, give rise to terms suppressed by more powers of $1/r$ and need not be considered when matching to the correlators of the monopole operator $O(\tau)$.

 \begin{figure}
    \centering
    \includegraphics[scale=0.25]{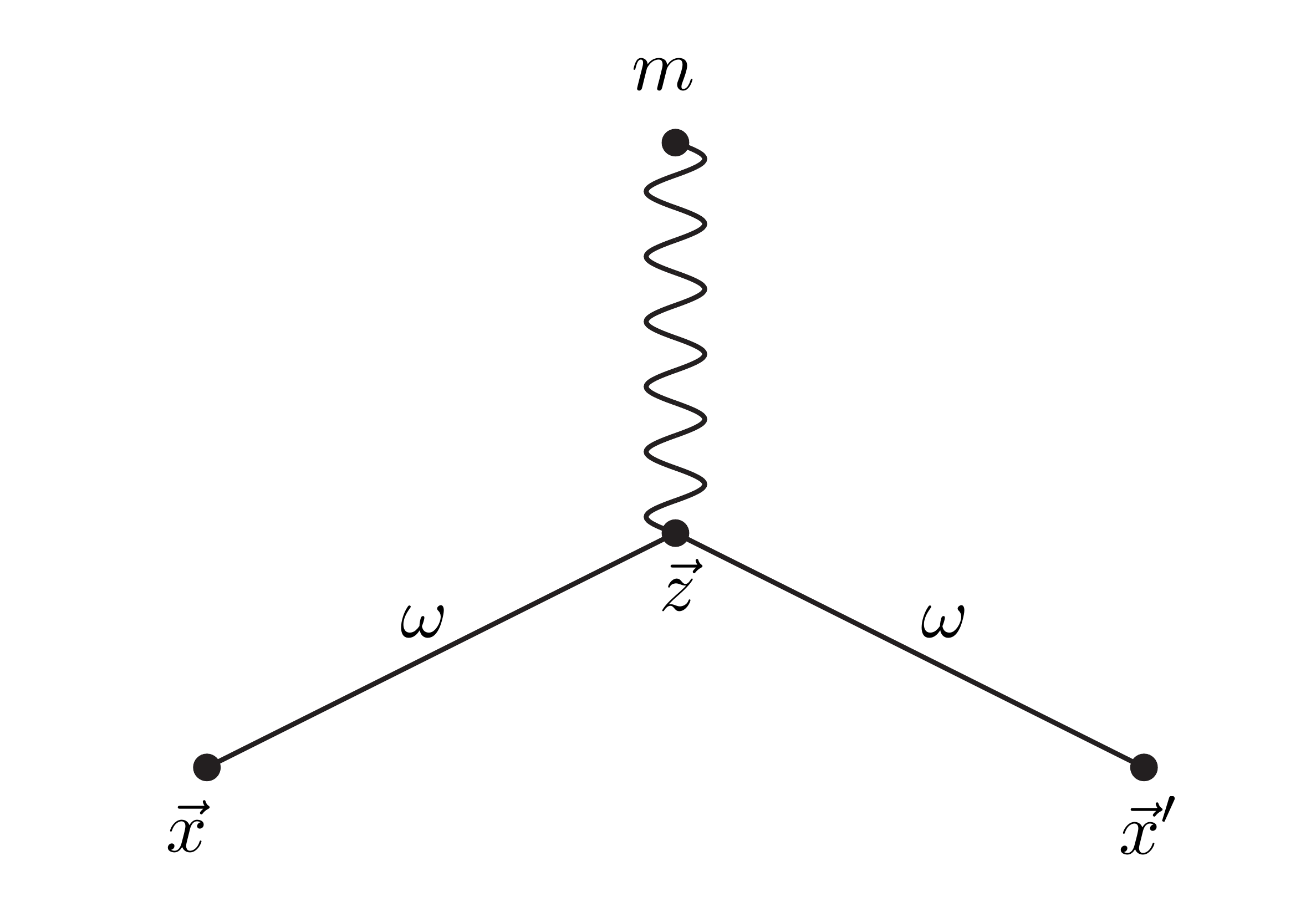}
\caption{Leading order potential graviton correction to bulk scalar Feynman propagator.}
\label{fig:pot}
\end{figure}

The potential exchange correction to the Feynman propagator is given in Fig~\ref{fig:pot}.   In order to evaluate the diagram, it is convenient to work in a mixed $(\omega,{\vec x})$ representation in which the free massless scalar Feynman propagator takes the form
\beq
D_F(\omega,{\vec x}) = -{i\over 4\pi |{\vec x}|} e^{i|\omega| |{\vec x}|},
\eeq
see also Eq.~(\ref{eq:fprop}).      In this representation the first perturbative correction to the scalar Feynman propagator  in the static background is
\beq
\label{eq:dg}
\Delta D_F(\omega,{\vec x},{\vec x}')=\mbox{Fig.}~\ref{fig:pot}= -{m \over 2 m_{Pl}^2} \omega^2 \int d^3{\vec z} D_F(0,{\vec z}) D_F(\omega,{\vec x}-{\vec z}) D_F(\omega,{\vec x}'-{\vec z}),
\eeq
where we have used the minimal gravitational coupling of the scalar to the potential graviton\footnote{We work in Feynman gauge where the graviton propagator tensor structure is $P_{\mu\nu,\alpha\beta}={1\over 2}\left[\eta_{\mu\alpha}\eta_{\nu\beta}+\eta_{\mu\beta}\eta_{\nu\alpha}-\eta_{\mu\nu}\eta_{\alpha\beta}\right]$.    The transformation between the radial coordinate defined by this choice of gauge and the Schwarzschild radial variable $r$ is given perturbatively to order $(r_s/ r)^3$ as $r\rightarrow\infty$ in ref.~\cite{GnR1}.}  ${\cal L}_{int} =-{1\over 2m_{Pl}} h^{\mu\nu} \left(\partial_\mu\phi\partial_\nu \phi - {1\over 2}\eta_{\mu\nu} (\partial\phi)^2\right)$.   Note that formally, this integral is log divergent in the IR, corresponding to ${\vec z}\rightarrow\infty$.    This is the usual IR singularity of scattering in a Coulomb field, which we regularize by giving the exchanged potential graviton a small mass $\mu$, so that the static propagator gets replaced by
\beq
D_F(0,{\vec x})=-{i\over 4\pi |{\vec x}|} \mapsto -{ie^{-\mu|{\vec x}|}\over 4\pi |{\vec x}|}. 
\eeq  

It is sufficient for our purposes to compute $\Delta G_F(\omega,{\vec x},{\vec x}')$ in the case ${\vec x}'=0$ and $|{\vec x}|\rightarrow\infty$.     It is trivial to do the angular integrals in Eq.~(\ref{eq:dg}), leaving over a radial integral that can be performed in terms of the incomplete Gamma function $\Gamma(0,z)$.  As $\mu\rightarrow 0,$
\beq
\Delta D_F(\omega,{\vec x},0)=  {r_s|\omega|\over 4\pi|{\vec x}|}\left[e^{i|\omega||{\vec x}|} \left(\gamma_E + \ln(\epsilon-2i|\omega| |{\vec x}|)\right) + e^{-i|\omega||{\vec x}|} \Gamma(0,\epsilon-2 i |\omega||{\vec x}|)\right],
\eeq
where $\epsilon = \mu |{\vec x}|\rightarrow 0^+$.  We have also used $r_s = 2 G_N m = m/16\pi m_{Pl}^2$.    In the limit $r\rightarrow\infty$ this has the asymptotic expansion
\beq
\Delta G_F(\omega,{\vec x},0)\sim  D_F(\omega,{\vec x}) \times r_s|\omega|\left({\pi\over 2} + i\ln (2|\omega||{\vec x}|)+ i\gamma_E \right) + {\cal O}(1/r^2).
\eeq
The form of the log in this expression is consistent with the well-known exponentiation of IR effects in the Coulomb propagator, i.e.
\beq
\label{eq:ddf}
D_F(\omega,r\rightarrow\infty,0)\sim -{ie^{i|\omega| r_*}\over 4 \pi r} \times \left[1+ {1\over 2} \pi r_s|\omega|\right],
\eeq
where the dependence on $r_* = r + r_s  \ln (2|\omega| r)$ accounts for the resummation of the leading IR singularity.    This reproduces the dependence of the mode functions on the Regge-Wheeler tortoise coordinate $r_*$ in the full theory as $r\rightarrow\infty$.   Terms suppressed by further powers of $1/r$ are gauge dependent, but not needed in our matching calculation.   On the other hand, the numerical factor of  $1+ {1\over 2} \pi r_s|\omega|$ multiplying the free propagator coincides with  the Sommerfeld factor associated with the Coulomb wavefunction at the origin, expanded in the limit $r_s\omega \ll 1$.

Given the correction to the Feynman propagator due to potential exchange, we immediately obtain the Dyson function as
\beq
\label{eq:ddd}
D_D(\omega,r,0)=D_F(-\omega,r,0)^*  \sim  {ie^{-i|\omega| r_*}\over 4 \pi r} \times \left[1+ {1\over 2} \pi r_s|\omega|\right].
\eeq
for $r\rightarrow\infty$.   Given that the identity between the Feynman, Dyson and Wightman functions
\beq
D_D(\omega,r,0)+D_F(-\omega,r,0)=W(\omega,r,0)+W(-\omega,r,0),
\eeq
and using the fact that the Wightman function is only supported for positive energy, we immediately obtain 
\beq
\label{eq:dw}
W(\omega,r\rightarrow\infty,0)\sim -{i\over 4\pi r} \theta(\omega) \left(e^{i\omega r_*}-e^{-i\omega r_*}\right)   \left[1+ {1\over 2} \pi r_s|\omega|\right].
\eeq

From these results, it is now straightforward to modify the matching calculation of sec.~\ref{sec:candelas} in order to account for potential exchange.   We simply have to replace the scalar propagators going from $x$ or $x'$ to the location of the source at ${\vec x}=0$ in Fig.~\ref{fig:candelas} by the Green's functions in Eqs~(\ref{eq:ddf})-(\ref{eq:dw}).    In the coincidence limit $r=r'\rightarrow\infty$, the rapidly oscillating phases $e^{\pm i \omega r_*}$ cancel as in the leading order calculation.  Thus the net effect of potential exchange is to multiply the integrand in Eq.~(\ref{eq:glo}) by the Sommerfeld factor $|\psi(0)|^2 \approx 1 + \pi r_s |\omega|$, which effectively replaces the worldline Wightman function  $A_+(\omega)$ by
\beq
\label{tail}
A_+(\omega)\rightarrow  A_+(\omega)\times (1 + \pi r_s |\omega|).
\eeq
 \begin{figure}
    \centering
    \includegraphics[scale=0.2]{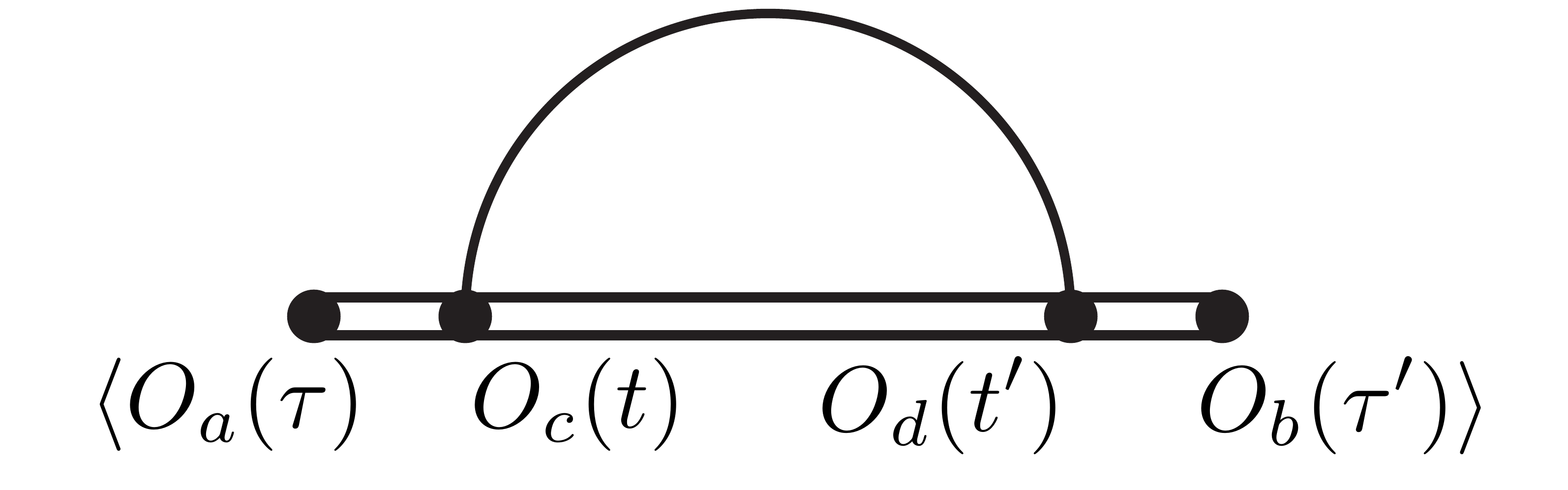}
\caption{Leading order correction to the in-in two-point functions $\langle O_a(\tau) O_b(\tau')\rangle$.}
\label{fig:4pt}
\end{figure}

To account for the corrections to the bulk Wightman function $\langle \phi(x) \phi(x')\rangle$ due higher insertions of the interaction $-\int d\tau \phi { O}$ in Fig.~\ref{fig:candelas}(c), we first determine how the interaction term $-\int d\tau \phi O$ modifies the in-in correlator matrix $\langle O_a(\tau) O_b(\tau')\rangle$, and then insert the result into the second term of Eq.~(\ref{eq:glo}).   The first correction to $\langle O_a(\tau) O_b(\tau')\rangle$  is represented in Fig.~\ref{fig:4pt}, which reads,
\beq
\label{eq:NLOoo}
\mbox{Fig.}~\ref{fig:4pt}= -\sum_{c,d}\int dt dt' \langle O_a(\tau) O_c(t) O_d(t') O_b(\tau')\rangle D_{cd}(t,t'),
\eeq
where $D(t,t')_{ab}$ is the matrix of free scalar propagators in Eq.~(\ref{eq:props}) with ${\vec x}={\vec x}'=0$.   To evaluate this expression, we use the result from sec.~\ref{sec:gauss} that the leading order correlators are Gaussian, of the form $\langle Q(t) Q(t')\rangle \approx 2 r_s \delta(t-t')$, and therefore
\beq
 \langle T O(t) O(t')\rangle = \langle {\tilde T} O(t) O(t')\rangle \approx 2 r_s \delta(t-t').
\eeq
It follows that all the four-point functions in Eq.~(\ref{eq:NLOoo}) are identical, and the diagram in Fig.~\ref{fig:4pt} is proportional to 
\beq
\sum_{cd} D_{cd}(x,x') = D_F(x-x') + D_D(x-x') - W_0(x-x') - W_0(x'-x) =0.
\eeq
Matching  Eq.~(\ref{eq:glo}) with the replacement Eq.~(\ref{tail}) to the full theory result Eq.~(\ref{eq:cunruh}) in the Unruh state Eq.~(\ref{full}), we find that to NLO,
\beq
\label{res}
A_+^U(\omega)=2\beta_H^{-1}\sigma_{abs}(|\omega|)\left[\theta(\omega)\left(1-{1\over 4} \beta_H\omega\right) + \theta(-\omega)\left(1+{3\over 4} \beta_H\omega\right)\right] + {\cal O}(r_s\omega)^3,
\eeq
and therefore $A^U_+(\omega)-A^U_+(-\omega) =\omega\sigma_{abs}(|\omega|) +  {\cal O}(r_s\omega)^3\approx 4\pi r_s^2\omega,$ which is in agreement with the results obtained in the Boulware vacuum, verifying the consistency relation Eq.~(\ref{eq:cr}) to leading non-trivial order in the power counting.

\section{Conclusions}

The effective field theory of quantum gravity in the low energy limit $\omega \ll M_{pl}$ in flat spacetime is well
understood. Calculations around non-trivial backgrounds are also under control, though technically more
nettlesome. However, black hole backgrounds present new conceptual challenges due to the existence Hawking radiation. Of course, for processes involving gravitationally
unperturbed backgrounds, such as the Hawking process itself, i.e. particle production, we have calculational
control. Now,  suppose we probe the black hole via some scattering process. Then we may
ask questions like, what is the induced change in the spectrum of Hawking radiation? Or what is the cross section
for the scattering of a charged particle off of a neutral black hole? To our knowledge these seem to be
open questions that have yet to be addressed.
 In this paper, we presented a formalism that can be used to answer these questions as well as others.    Our framework captures the universal long distance effects of both classical and quantum horizon dynamics of black holes, extending the EFT approach to low energy quantum gravity beyond the canonical graviton loop effects that have been traditionally studied.

In the EFT the ``internal dynamics'' of the black hole are captured by a set of correlation functions of composite operators that live on the worldline.
In  \cite{GnR2}  we used such ideas to describe classical absorptive processes, which entailed extracting the
Wightman function $A_+(\omega>0)$.  Emission processes on the other hand are described by $A_+(\omega<0)$.
Detailed balance dictates that $A_+(\pm\omega)$ must have the same scaling in $m_{Pl}$ which at first may seem
surprising given that Hawking radiation is inherently quantum mechanical. However, this lack of suppression is a consequence of  the well known fact that the Hawking rate $d\Gamma/d\omega$ has no factors of $\hbar$.  Here we have shown that for scalar fields coupling to a black hole, at next to leading order in the derivative expansion,
the correlators are given by Eq.~(\ref{res}). The ensuing retarded causal Green's function responsible for all classical observables, such as
the black hole Love number, has only Planck suppressed contributions, as expected, since in free field theory the state dependence must be
absent.

While in this paper we only considered the simplest toy model of a scalar field coupled to the BH horizon, most of the observations and results of this paper carry over straightforwardly to a low energy EFT of gauge, fermion, or graviton fields interacting with the BH.    We also expect similar results to hold for spinning and/or charged black holes\footnote{Indeed, as pointed out in~\cite{wald}, the transition probabilities $p(m\rightarrow n)$ obtained in~\cite{bekenstein,wald} are also valid for spinning black holes emitting modes in the super-radiant regime.  Even though the results of~\cite{bekenstein,wald} were strictly speaking obtained for the case of non-interacting spin-0 particles only, the derivation can be generalized straightforwardly to account for higher spins as well.}.  For instance, in the case of gravity~\cite{GnR2}, the BH couples linearly to the Weyl tensor, and Eq.~(\ref{eq:EFT}) generalizes to
\begin{equation}
S_{int} = -\int d\tau \left(Q^E_{\mu\nu} E^{\mu\nu}+Q^B_{\mu\nu} B^{\mu\nu}\right)+\cdots,
\end{equation}
where the electric quadrupole operator $Q^E_{\mu\nu}$ couples to $E_{\mu\nu} = W_{\mu\rho\nu\sigma} {\dot x}^\rho {\dot x}^\sigma$, the magnetic moment operator $Q^B_{\mu\nu}$ to the dual, $B_{\mu\nu} = {\tilde W}_{\mu\rho\nu\sigma} {\dot x}^\rho {\dot x}^\sigma$, and higher moments, coupled to gradients of the Weyl tensor, are not shown.    Similarly, the long wavelength dynamics of the photon is well approximated by electric and magnetic dipole operators whose correlators can be matched to the full theory by the methods described in this paper.

 \begin{figure}
    \centering
    \includegraphics[scale=0.2]{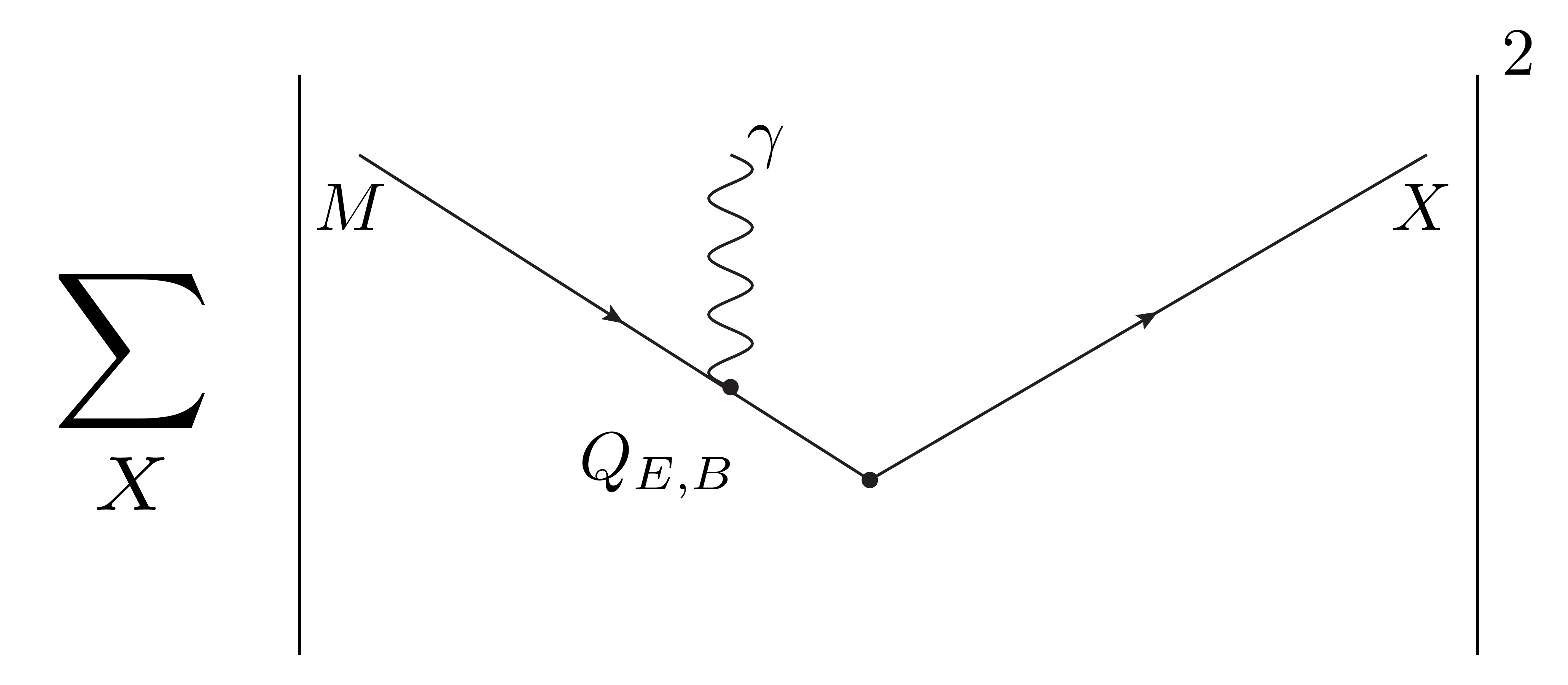}
\label{fig:soft}
\caption{The diagram responsible for inclusive photon production induced by the deflection of a black hole due
to the interactions with an external current. The soft emission factorizes from the hard scattering off of the source.}
\end{figure}

As an example of the types of observables that are in principle accessible to our formalism, it is possible to formulate a ``soft photon'' theorem, analogous to the standard result in QED~\cite{Weinberg:1965nx}, that expresses the factorization of soft Hawking radiation from an \emph{electrically neutral} BH that participates in an otherwise arbitrary scattering process.     Unlike the case of QED, the soft theorem in our effective theory is formulated  at the level of the squared amplitude, where the final states of
the black hole  $|X\rangle$  are summed over (see Fig.~\ref{fig:soft}).    However, like the QED result, the BH soft theorem encapsulates the universal aspects of the emitted electromagnetic radiation in the limit in which the outgoing Hawking photons have energy $E_\gamma\rightarrow 0$.   Finally, our EFT has the predictive power to calculate effects of quantum corrections mediated by off-shell Hawking exchange  to various processes such as,  inelastic scattering cross sections involving one or more black holes as asymptotic scales, and QED corrections to energy level of charged particles bound to black holes.
These results will be presented in forthcoming publications.

\section{Acknowledgments}

We are supported by DOE HEP grants DE-SC00-17660 (WG) and DE- FG02-04ER41338 and FG02-06ER41449 (IZR).   WG thanks the UCSD theory group for financial support while this work was being completed.    We thank  Simon Caron-Huot, Aneesh Manohar, and Edward Witten for helpful discussions.


\end{document}